\documentclass[aps,nofootinbib,superscriptaddress,twocolumn,floatfix]{revtex4}
\usepackage{amsmath}
\usepackage{latexsym}
\usepackage{graphicx}
\usepackage{color}
\usepackage{fancyhdr}
\usepackage{hyperref}

\begin{document}


\title{Quantum entanglement distribution with 810 nm photons through
telecom fibers}

\author{E. Meyer-Scott}

\affiliation{Institute for Quantum Computing, University of Waterloo, 200 University Avenue W, Waterloo
N2L 3G1, Canada}
\email{emeyersc@iqc.ca}
\author{H. H\"{u}bel}
\affiliation{Institute for Quantum Computing, University of Waterloo, 200 University Avenue W, Waterloo
N2L 3G1, Canada}
\author{A. Fedrizzi}
\affiliation{Department of Physics and Centre for Quantum Computer Technology, University of Queensland, Brisbane
QLD 4072, Australia}
\author{C. Erven}
\affiliation{Institute for Quantum Computing, University of Waterloo, 200 University Avenue W, Waterloo
N2L 3G1, Canada}
\author{G. Weihs}
\affiliation{Institute f\"{u}r Experimentalphysik, Universit\"at Innsbruck, Technikerstrasse 25, 6020 Innsbruck, Austria}
\affiliation{Institute for Quantum Computing, University of Waterloo, 200 University Avenue W, Waterloo
N2L 3G1, Canada}
\author{T. Jennewein}
\affiliation{Institute for Quantum Computing, University of Waterloo, 200 University Avenue W, Waterloo
N2L 3G1, Canada}

\begin{abstract}
\noindent We demonstrate the distribution of polarization entangled photons of wavelength 810 nm through
standard telecom fibers. This technique allows quantum communication protocols to be performed
over established fiber infrastructure, and makes use of the smaller and better performing setups
available around 800 nm, as compared to those which use telecom wavelengths around 1550 nm.
We examine the excitation and subsequent quenching of higher-order spatial modes in telecom
fibers up to 6 km in length, and perform a distribution of high quality entanglement (visibility
95.6$\%$). Finally, we demonstrate quantum key distribution using entangled 810 nm photons over a
4.4 km long installed telecom fiber link.
\end{abstract}

\maketitle

The ability to distribute entanglement is an important
building block in the field of quantum information processing.
It is employed in protocols such as quantum
teleportation\cite{Bouwmeester1997Experime}, quantum key distribution (QKD)\cite{RevModPhys.74.145}, and quantum
computing \cite{Knill2001A-scheme}. Many quantum information experiments so
far have been performed at wavelengths around 800 nm,
making use of the high performance (low noise, high speed,
and around 70$\%$ efficiency) of silicon avalanche photodiodes
(Si-APDs) for single-photon detection. Such systems have
been demonstrated in laboratories, over free space links \cite{Kurtsiefer2002Quantum-,ursin-2007-3} or
with custom laid 800 nm single mode fibers\cite{Poppe:04}, the latter of
which face the difficulty of installing a dedicated link. In
addition, most quantum memory implementations and
quantum dot photon-sources are designed around 800 nm
optical transitions\cite{Yuan2008Experime}. On the other hand, quantum communication
setups have been built to make use of existing telecom
fiber infrastructure and low fiber loss at 1550 nm\cite{Hubel:07}; however,
single-photon detectors based on APDs designed for these
wavelengths (InGaAs-APDs) add considerable complexity,
require elaborate synchronization of detector gates, and suffer
from low detector efficiencies ($\sim$15$\%$). In fact, based on
the efficiencies above, and fiber losses of 3 dB/km for 800
nm light and 0.22 dB/km for 1550 nm light, overall attenuation
will be lower for 800 nm photons for up to 7.3 dB of
fiber losses, corresponding to 2.4 km of telecom fiber. In
spite of the common perception that entanglement distribution
at 800 nm strictly requires wavelength-specific components,
it is obvious that such short wavelength systems
would greatly benefit if used with the existing fiber infrastructure.
Toward this goal we report on the high fidelity
distribution of entangled photon pairs at 810 nm through
several kilometers of standard telecom fibers, which provides
a path for demonstrating quantum information applications,
like entanglement based QKD, and other quantum optics experiments
in existing fiber networks.

We adapted a polarization-entanglement based quantum
communication system\cite{Erven2008Entangle} at 810 nm using the BBM92
protocol\cite{Bennett1992Quantum-} to transmit photons to Alice and Bob through
varying lengths of single mode telecom fibers [mode field
diameter (MFD) of 9.2 $\mu$m] ranging from 250 m to 6000 m
(Fig. 1) or short stretches of 810 nm single mode fibers with
a MFD of 5.4 $\mu$m. After transmission through the fibers,
each photon of the entangled pair passes through a polarization
analyzing module, which forwards the photons to one of
four Si-APDs, depending on the measured polarization state
(0$^{\circ}$, 90$^{\circ}$, 45$^{\circ}$ or -45$^{\circ}$). A time-tagging unit then records the
state and the time of the detection. This information is
bundled and passed to Alice and Bob's computers, which
communicate classically to find the optimal time offset to
maximize the number of coincidences between Alice's and
Bob's detection events.

\begin{figure}[b]
\begin{center}
\includegraphics[width=3in]{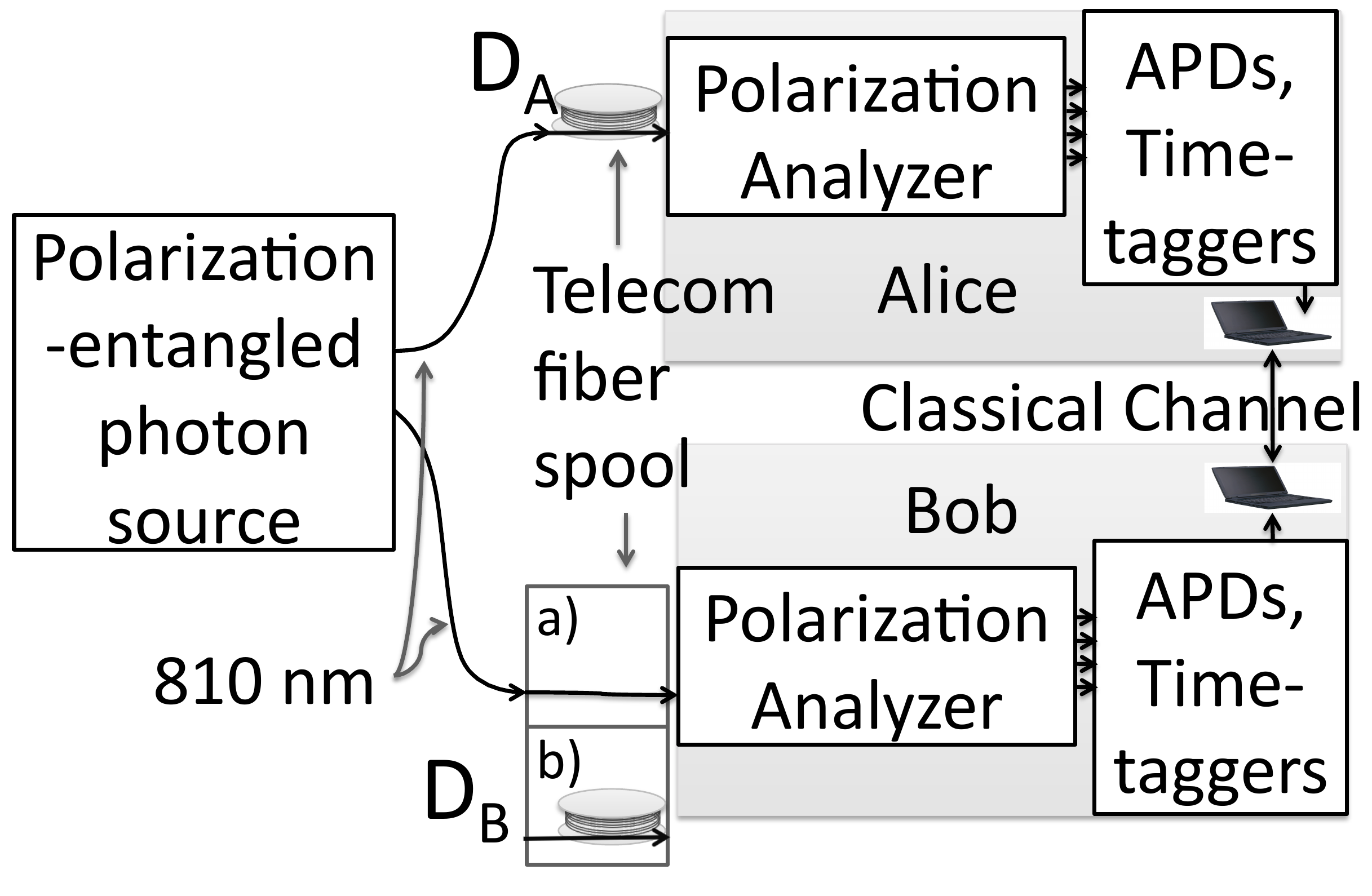}
\caption{Schematic of setup. (a) Asymmetric distribution scheme, with only
a short 810 nm single mode fiber to Bob. (b) Symmetric scheme, with long
1550 nm telecom fibers to both Alice (length D$_{A}$) and Bob (length D$_{B}$)}
\end{center}
\end{figure}

Since standard telecom fiber is slightly multimode for
810 nm light we expect the appearance of higher-order
spatial modes. Guided wave theory predicts two linearly polarized
modes of propagation (LP$_{01}$ and LP$_{11}$) \cite{Saleh2007Fundamen} for
810 nm photons in a telecom fiber. Here the modes are labeled
based on the distribution arm (Alice or Bob) and the
azimuthal index i (e.g., A$_{i1}$). The two propagation modes show modal dispersion; i.e., the group velocity of the A$_{11}$
mode is different from that of the A$_{01}$ mode, resulting in two
distinct arrival times \cite{Townsend1998Experime}. Detecting Bob's photons locally and
Alice's after 3 km of telecom fiber resulted in a histogram of
coincidences with two pronounced peaks, as seen in Fig. 2.
The relative offset of the two peaks varied linearly with fiber
length leading to a measured modal dispersion of 2.20 ns/
km, in excellent agreement with the theoretical value of 2.19
ns/km. As evidenced by the well-defined peaks in Fig. 2,
there is little crosstalk between the two modes after the initial
insertion, so the polarization state in the fundamental
mode is well preserved and the timing signature of each
mode is evident.

\begin{figure}[t]
\begin{center}
\includegraphics[width=3in]{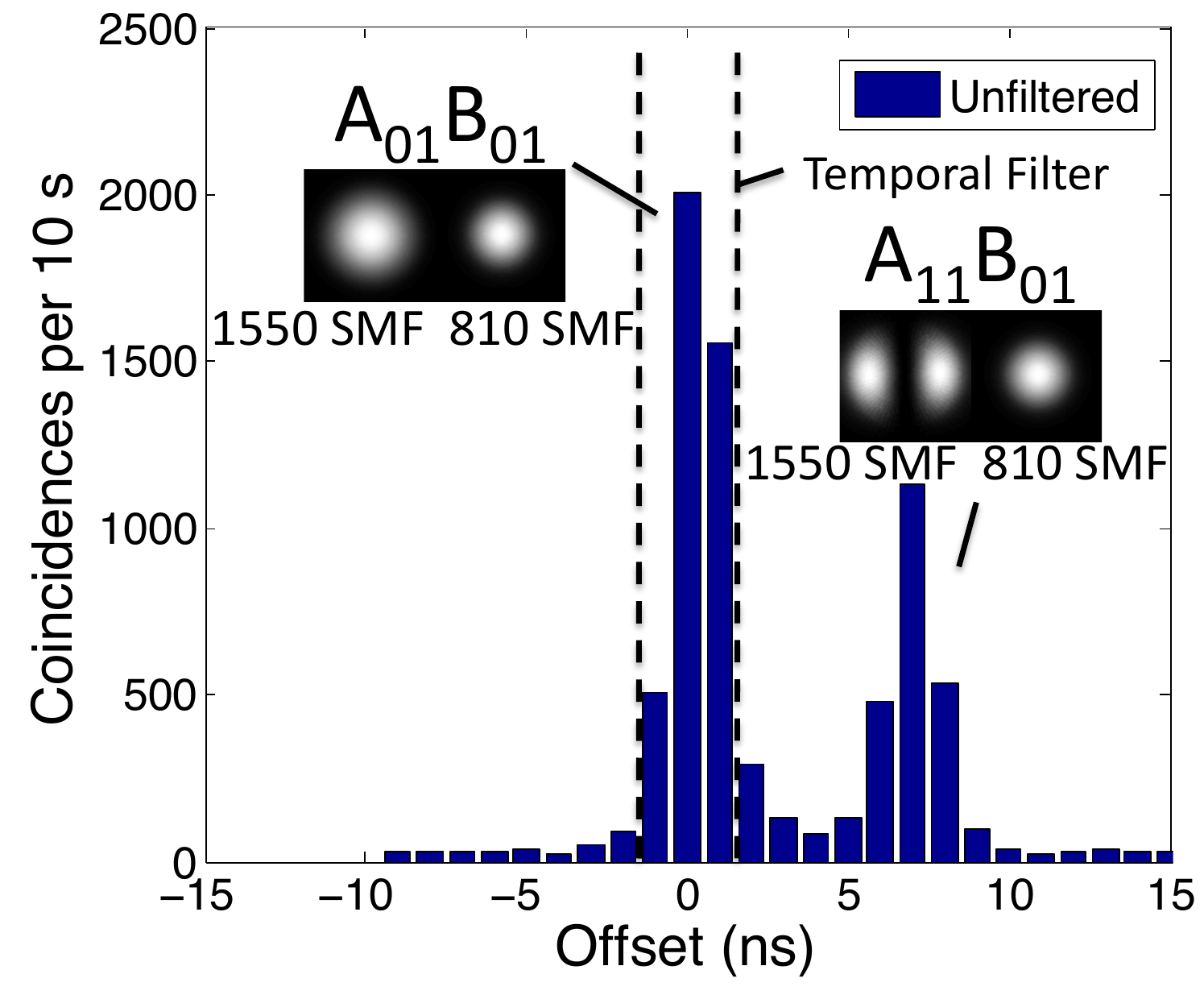}
\caption{Asymmetric distribution (D$_{A}$=3 km, D$_{B}$=0 km):
histogram of coincident detection events with delay between Alice's and
Bob's detection. The slower A$_{11}$ mode in Alice's arm is intentionally excited to illustrate the effect, then filtered. The theoretical power distribution of the propagating spatial modes at 810 nm for different fibers types is inset.}
\end{center}
\end{figure}

For a high fidelity transmission of polarization entangled
photons it is necessary to select only the fundamental mode
in both arms (A$_{01}$B$_{01}$), as higher order modes will lead to an
increased error in the polarization contrast since only one of
the polarization rotations experienced by different modes in
the fiber can be compensated for. In principle, the modes
could be separated and compensated individually, but without
such elaborate mode extraction two methods for filtering
out the higher modes at the receiver are developed:

(i) 
\hspace{4 pt} In the case of an asymmetric distribution, where the
fiber lengths to Alice and Bob are not identical, a
temporal filter can be applied in the form a of a narrow
coincidence window, which will cut out the
higher order peak, as demonstrated in Fig. 2. This technique introduces no additional optical losses.

(ii)\hspace{4 pt}  In a symmetric distribution, where both Alice and Bob
receive photons through telecom fiber and where the
difference in fiber lengths is less than 2 km, the
A$_{01}$B$_{01}$ and A$_{11}$B$_{11}$ peaks become inseparable in time,
as seen in the central peak of Fig. 3. In this case, in
addition to the temporal filter to eliminate the side
peaks, a spatial filter (810 nm single mode fiber) is
used before detection. Since the radial extent of the
power in the higher order modes is greater than that in
the fundamental mode, the smaller core of the 810 nm
fiber (shown as a ring in Fig. 3) removes around 98$\%$
of the A$_{11}$ or B$_{11}$ mode while preserving at least 75$\%$
of the A$_{01}$ or B$_{01}$ mode.

\begin{figure}[t]
\begin{center}
\includegraphics[width=3in]{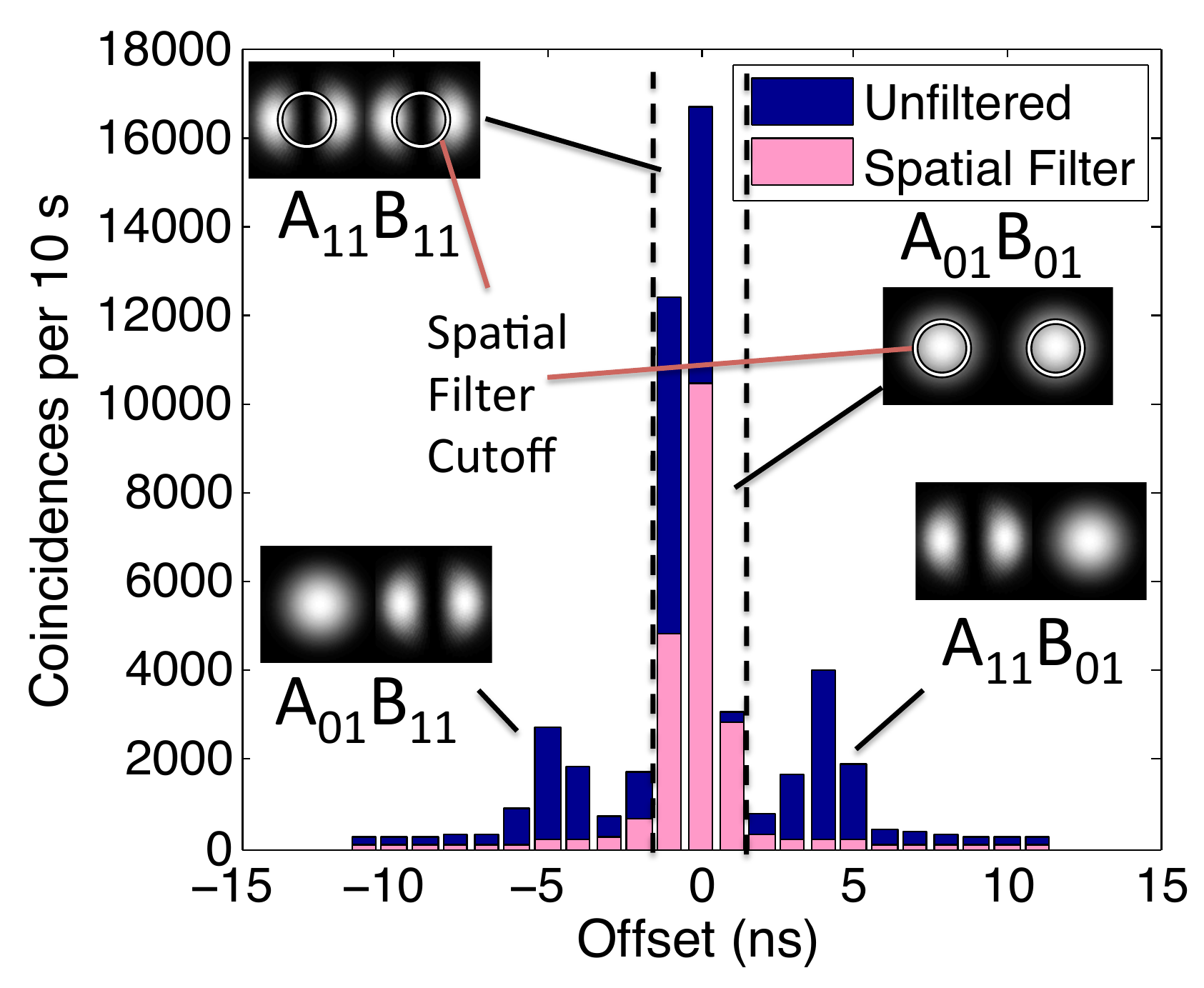}
\caption{Symmetric distribution (D$_{A}$=2 km,D$_{B}$=2 km): selection
of A$_{01}$B$_{01}$ mode by spatial filtering. The spatial filter (core radius
superimposed on power distribution) eliminates not only the side peaks (due
to photons coupling into cross modes) but also those coincidences in the
central peak that are a result of both photons coupling into the higher mode
(A$_{11}$B$_{11}$), such that they give little or no betraying time offset.}
\end{center}
\end{figure}

In order to quantify our filtering methods we performed
entanglement distribution measurements with telecom fiber
spools of lengths up to D$_{A}$=6 km (asymmetrically), and up
to D$_{A}$=D$_{B}$=2 km (symmetrically). We extracted the entanglement
visibility (a measure of the quality of
entanglement)\cite{Hubel:07}, raw coincidences, and secure key rate for
QKD based on realistic error correction and privacy
amplification\cite{2007PhRvA..76a2307M} (Table I). To set a benchmark for comparison,
measurements were performed locally with short 810 nm fibers (2 m), resulting 
in 95.7$\pm$0.4$\%$ visibility\footnote[1]{Differences in count rates are due to a realignment of the source between
measurements.}.

\begin{table*}[hc]
\caption{Summary of entanglement distribution for various telecom fiber lengths, including QKD key rates. Transmission loss includes attenuation in the
optical fibers, as well as loss from the filtering processes. Local measurements gave an average visibility of 95.7$\pm$0.4$\%$. Uncertainty is taken as due to
Poissonian count fluctuations.}
\begin{tabular}{l c c c c c c c}
\hline
Distribution 	& D$_{A}$ (km) & D$_{B}$ (km) & Transmission & Filtering & Visibility  		& Coinc. & Secure Key \\
Scheme		&			&			& Loss (dB)	&		&	(\%)			&(rate/s)& (rate/s)\\
\hline
Asymmetric & 2.0 & 0.0 & 6 & None & 88.0$\pm$0.2 & 3000 & 420 \\
& & & 7 & Temporal & 94.6$\pm$0.2 & 2700 & 800 \\
Asymmetric & 5.0 & 0.0 & 15 & Temporal & 91.6$\pm$0.5 & 430 & 90 \\
Symmetric & 2.0 & 2.0 & 12 & None & 62.9$\pm$0.4 & 5200 & 0 \\
& & & 14 & Temporal & 92.2$\pm$0.3 & 3600 & 850 \\
& & & 16 & Temporal+spatial & 95.6$\pm$0.2 & 1950 & 650 \\
\hline
\end{tabular}

\end{table*}

For the asymmetric distribution we employed a 3 ns coincidence
window as the temporal filter: at 2 km of fiber in one arm, for example, overall visibility was improved from
88.0$\%$ to 94.6$\%$ with this method. Figure 4 shows the visibility
for asymmetric distribution distances up to 6 km. Figure
4 also shows a calculated QKD secure key rate, dropping
linearly (on the log scale) with increasing lengths of fibers,
with a sharp cut-off around 6 km due to increased loss and
detector dark counts.

\begin{figure}[t]
\begin{center}
\includegraphics[width=3in]{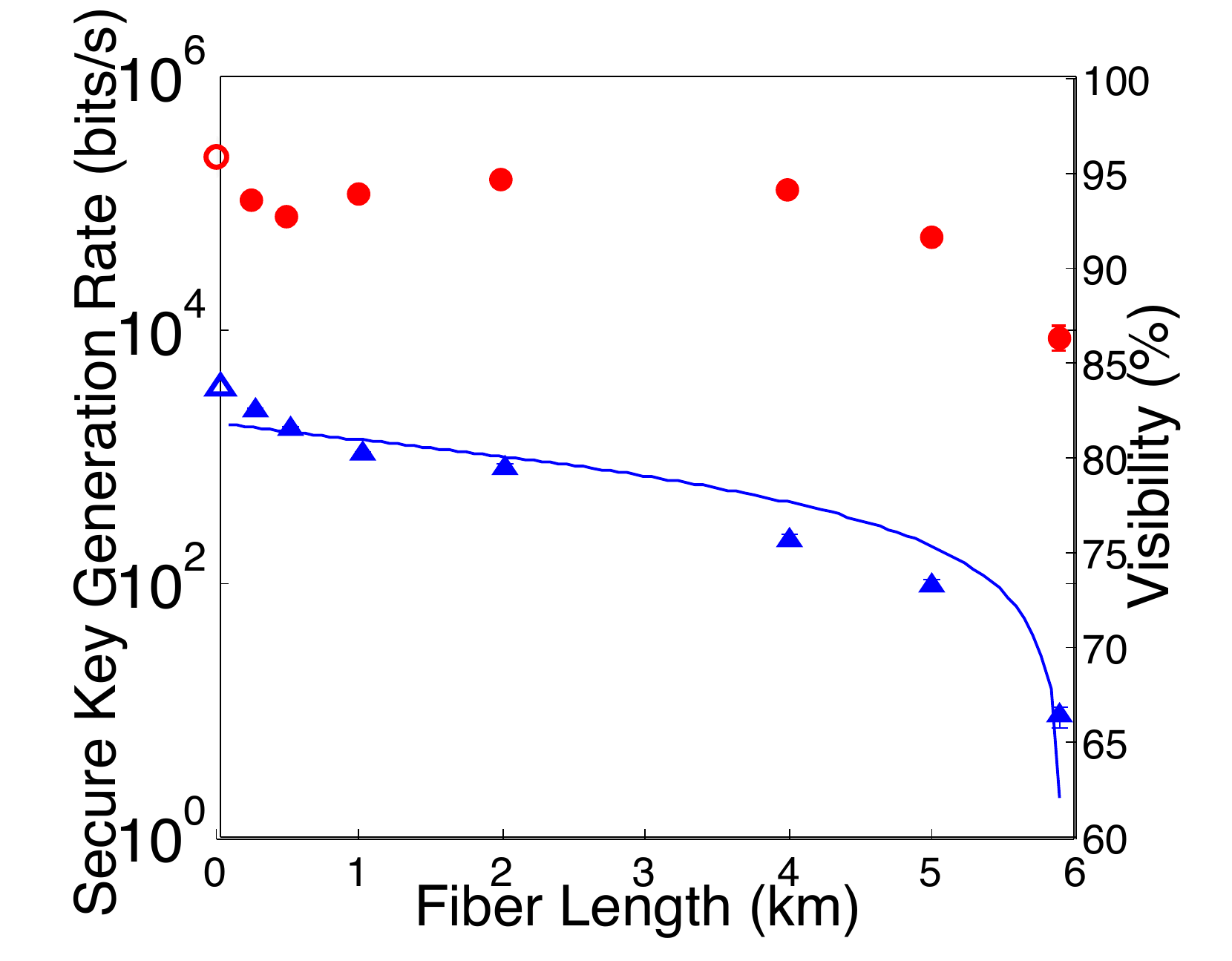}
\caption{Secure key generation rates (triangles) and measured
visibility (circles) as a function of fiber length in Alice's arm. The open
symbols are local data, and the line is a fit to the secure key rate with
realistic parameters (see Ref. 13). Error bars are smaller than symbol size.}
\end{center}
\end{figure}

In the case of symmetric distribution (see Table I), we
employed both filtering techniques to raise visibility from
62$\%$ to 95.6$\%$. For this and the asymmetric case, visibility is
brought close to the benchmark which implies that the higher
order modes are suppressed using the filtering detailed
above, and that there is no significant crosstalk along the
length of the fiber.

To illustrate the utility of this form of entanglement distribution,
we performed a full QKD protocol over two symmetric
2.2 km channels of installed telecom fibers. Two
parallel fibers were used between the Mathematics and Computer
Building on the University of Waterloo's campus
(source of entangled photons) and the Perimeter Institute
(detection modules Alice and Bob), leading to a total distribution
distance of 4.4 km. The quantum bit error ratio
(QBER) was higher than for the fiber spools, likely due to
disturbances from passing cars, trains, and thermal fluctuations.
For example, over 15 min, the average QBER was
4.3$\%$ (i.e., 91.4$\%$ visibility) with both temporal and spatial
filtering, leading to an average secure key rate of 350 bits/s (see Fig. 5). During longer runs, the errors tended to increase
with time due to polarization drifts in the fibers.

\begin{figure}[htp]
\begin{center}
\includegraphics[width=3in]{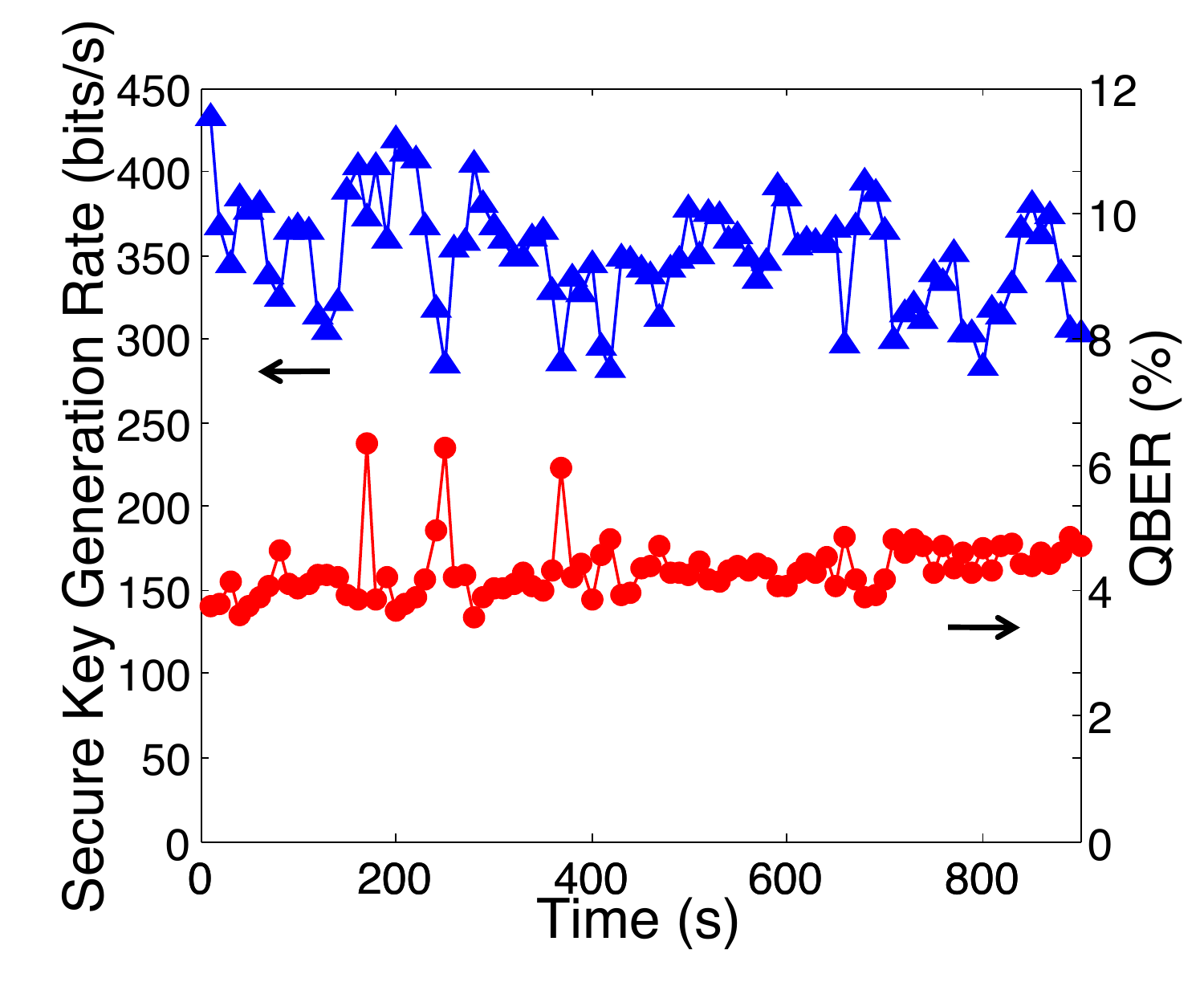}
\caption{Secure key generation rate and measured QBER in a
symmetric installed fiber link. Each data point is the average over 20 s.}
\end{center}
\end{figure}

Our approach has the potential for much higher distribution
rates, as local coincidence rates of 2.5 MHz have been
achieved with short-wavelength entanglement sources \cite{1367-2630-11-8-085002}. Assuming
the same fiber loss observed here, secure key rates of
500 kb/s are possible over a 4 km symmetric link using only
standard telecom fibers. Previous systems based on weak laser
pulses around 800 nm were limited to key rates of 100
kb/s over 4 km of telecom fiber \cite{collins:180}. Given the superior functionality
and lower complexity of detectors at 800 nm combined
with the multiuser networking capabilities of
entanglement \cite{Lim2008Distribu}, we believe that such QKD systems will find
applications in inner city links or corporate networks. In addition,
the possibility to address multiple modes in a fiber
could be useful for implementations of higher-dimensional
quantum information.

We have demonstrated the viability of entanglement distribution
in standard optical fibers using 810 nm photons.
With suitable filtering, error rates are not affected by higher
order modes in the fiber and high fidelity distribution can be
achieved over several kilometers. We believe our results
pave the way for a wide usage of telecom optical infrastructure
together with the well established quantum information
systems at shorter wavelengths.

The authors would like to thank Allison MacDonald for
calculations, Greg Cummings and Bruce Campbell of IST
UW, John McCormick, Joy Montgomery, Hon Lau, and Joe
Stauttener of IT at PI, Ch\^{a}teauneuf Fran\c{c}ois of INO for test
fibers, NSERC (CGS, PGS, QuantumWorks, Discovery),
OCE, CIFAR, CFI, and ERA for funding.
\def\urlprefix{}
   \def\url#1{}

\bibliography{/Users/Evan/Documents/IQC/IQC}
\bibliographystyle{apsrev}

\end{document}